\documentclass[conference]{IEEEtran}
\IEEEoverridecommandlockouts
\usepackage{cite}
\usepackage{amsmath,amssymb,amsfonts}
\usepackage{algorithmic}
\usepackage{graphicx}
\usepackage{textcomp}
\usepackage{xcolor}
\usepackage{hyperref}
\usepackage{multirow}
\usepackage{rotating}
\usepackage{subfigure}
\usepackage{float}
\usepackage{adjustbox}

\def\BibTeX{{\rm B\kern-.05em{\sc i\kern-.025em b}\kern-.08em
    T\kern-.1667em\lower.7ex\hbox{E}\kern-.125emX}}
\begin{document}

\title{ELiOT : End-to-end Lidar Odometry using Transformer Framework 
\thanks{\textsuperscript{1} Department of Electrical Engineering, Korea Advanced Institute of Science and Technologies (KAIST), Daejeon, Republic of Korea
        {\texttt{\{lee.dk, hcshim\}@kaist.ac.kr}}}
\thanks{\textsuperscript{2} Robotics Program, KAIST, Daejeon, Republic of Korea
        {\texttt{menu107@kaist.ac.kr}}
        }
\thanks{*Corresponding Author.}
\thanks{This work was supported by the Technology Innovation Program (RS-2023-00256794, Development of drone-robot cooperative multimodal delivery technology for cargo with a maximum weight of 40kg in urban areas) funded By the Ministry of Trade, Industry \& Energy(MOTIE, Korea).}
\thanks{This work has been submitted to the IEEE for possible publication. Copyright may be transferred without notice, after which this version may no longer be accessible.}
}

\makeatletter
\newcommand{\linebreakand}{%
  \end{@IEEEauthorhalign}
  \hfill\mbox{}\par
  \mbox{}\hfill\begin{@IEEEauthorhalign}
}
\makeatother

\author{\IEEEauthorblockN{Daegyu Lee\textsuperscript{1}}
\and
\IEEEauthorblockN{Hyunwoo Nam\textsuperscript{2}}
\and
\IEEEauthorblockN{D.Hyunchul Shim\textsuperscript{1*}}
}

\maketitle


\begin{abstract}
In recent years, deep-learning-based point cloud registration methods have shown significant promise. 
Furthermore, learning-based 3D detectors have demonstrated their effectiveness in encoding semantic information from LiDAR data. 
In this paper, we introduce \textbf{ELiOT}, an end-to-end LiDAR odometry framework built on a transformer architecture. 
Our proposed Self-attention flow embedding network implicitly represents the motion of sequential LiDAR scenes, bypassing the need for 3D-2D projections traditionally used in such tasks. 
The network pipeline, composed of a 3D transformer encoder-decoder, has shown effectiveness in predicting poses on urban datasets. 
In terms of translational and rotational errors, our proposed method yields encouraging results, with 7.59$\%$ and 2.67$\%$ respectively on the KITTI odometry dataset. 
Our end-to-end approach forgoes conventional geometric concepts and leverages positional embedding networks to identify motion flow and emphasize key features, achieving geometric motion in a seamless manner.
\end{abstract}

\begin{IEEEkeywords}
Localization, LiDAR odometry, Deep learning
\end{IEEEkeywords}

\section{Introduction}
Localization is one of the most essential modules in autonomous mobile robot. In recent years, LiDAR (Light Detection and Ranging)-based and vision-based \cite{qin2018vins, cao2022gvins, cvivsic2022soft2} localization methods have gained substantial interest because the localization quality of global navigation satellite system (GNSS) deteriorates in urban canyons due to the signal multi-path and weak signal strength \cite{kamat2018survey, sun2020robust}.
Especially, point-clouds obtained from a LiDAR or a RGB-D camera can be utilized to estimate the robot motion between two consecutive sweeps.
Therefore, this application called the LiDAR odometry (LO) \cite{jonnavithula2021lidar} is studied using geometry-based method \cite{zhang2014loam, shan2018lego, shan2020lio} or learning-based method \cite{yin2018locnet, lu2019l3, barsan2020learning, horn2020deepclr}. 
\par
The goal of the LO is to minimize the error of translation and rotation while accumulating the sequential robot motions. Typically, traditional LO methods are based on the point registration, which iteratively finds  point-to-geometry correspondences, and minimizes the pose transformation error.
Iterative Closest Point (ICP) \cite{besl1992method}, G-ICP \cite{segal2009generalized}, Normal Distributions Transform (NDT) \cite{biber2003normal} and its variants \cite{censi2008icp, das2012scan, das2014scan} are well-known examples utilizing the point registration algorithm. 
Rather than employing computationally expensive dense registration methods, feature-based LO utilizes efficient geometry constraint calculations for its implementation as exemplified in the previous works \cite{shan2020lio, shan2018lego,xu2021fast} 
This method has shown high performance on various benchmarks including the KITTI dataset \cite{geiger2012we}.
Despite these advancements, geometry-based methods may not adequately account for memory considerations, whereas learning-based methods leverage data-driven approaches, as demonstrated in the work of Xue \textit{et al} \cite{xue2020deep}.
\par
In addition, learning-based approaches have a strength in handling the point-cloud acquired in deteriorated conditions because deep learning for feature extraction is more promising than handcrafted feature extractions by modeling robust and effective cross-domain generalizability as in \cite{almalioglu2022deep}.
Recent state-of-the-art learning-based LiDAR odometry methods harness the advantages of a cylindrical projection-based approach \cite{wang2021efficient, liu2023translo}. This form of projection representation has the benefit of allowing the use of image-based network pipelines and concepts from recent architectural studies.
However, we note that emerging 3D data detection studies indicates that 2D represented methods cannot fully reflect the characteristics of 3D data and easy to loose features.
\par
\begin{figure}[!t]
    \centering
    \includegraphics[width=1.0\columnwidth]{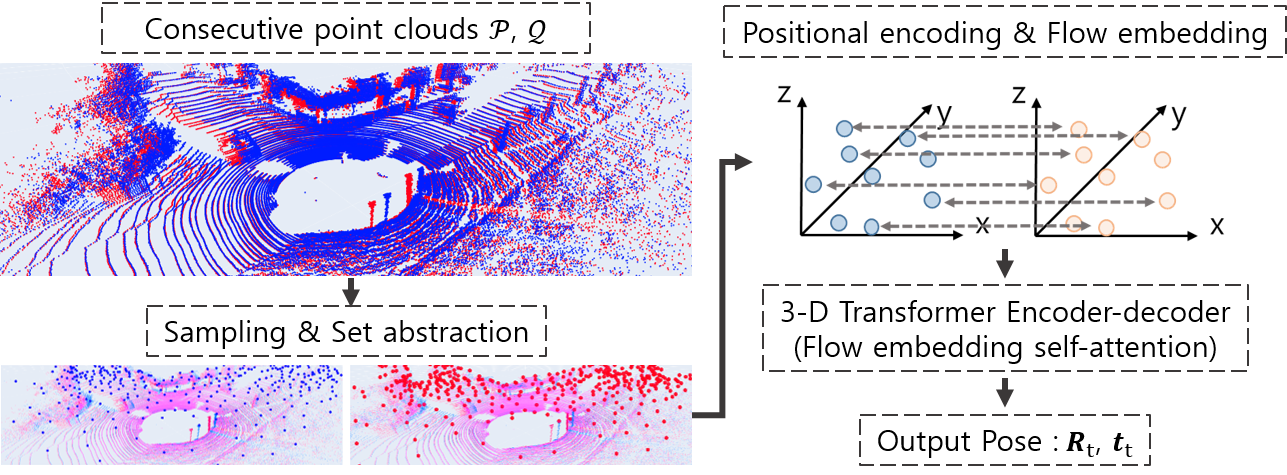}
    \caption[Propose end-to-end LO using 3-D Transformer frameworks, \textbf{ELiOT}.]{Propose end-to-end LO using 3-D Transformer frameworks, \textbf{ELiOT}.}
    \label{fig:simple}
\end{figure}
In this work, we propose end-to-end LO using 3-D Transformer frameworks, \textbf{ELiOT}, as illustrated in Fig. \ref{fig:simple}. 
The main contributions of this paper is as follows.
\begin{itemize}
    \item We present an end-to-end approach that foregoes the need for conventional geometric concepts such as K-nearest neighbors (\textit{KNN}) and clustering algorithms.
    \item We introduce a self-attention flow embedding network designed to implicitly represent the motion within sequential LiDAR scenes.
    \item To the best of our knowledge, we are the first to propose a 3D transformer-based LiDAR Odometry (LO) framework that eliminates the need for 3D-2D projection.
\end{itemize}
The remainder of this paper is organized as follows. Section \ref{sec:related} introduces related 3D object detectors and deep LO studies. 
Section \ref{sec:methods} describes our problem definition, network and its modules for LO.
Section \ref{sec:implement} describes our implementation details.
Experiments conducted on the KITTI odometry dataset and our high-speed racecar dataset are discussed in Section \ref{sec:experiment}. 
Finally, Section \ref{sec:conclusion} concludes this study.

\section{Related work}
\label{sec:related}
\subsection{Feature extraction of 3-D detector}
In recent years, 3-D detectors are studied to handle the sparse and unordered characteristic of points.
Before directly processing point clouds in 3D, there has been a lot of research that has rasterized data into 2D by converting it to a BEV (Bird's Eye View) in order to utilize image-based recognition models for object recognition \cite{Yang_2018_CVPR, lang2019pointpillars}. 
This method of representing 3D data in 2D does not fully reflect the characteristics of 3D data and has difficulty considering the sparse data characteristics. As a result, research has emerged to directly process sparse, unordered set of 3D points as in \cite{qi2017pointnet, qi2017pointnet++}.
\par
However, because processing data arrays with tens of millions of points requires a large amount of computation, research has been proposed to efficiently detect objects by voxelizing the 3D space \cite{zhou2018voxelnet, deng2021voxel}. 
However, this voxelization method can decrease the accuracy of recognizing small objects, leading to the emergence of research that combines the benefits of both point-wise and voxel-wise approaches through a hybrid method of point-voxel feature extraction \cite{shi2020pv, shi2022pv}.
\par
In addition, the Transformer architecture by Vaswani \textit{et al.} \cite{vaswani2017attention} has been successfully adopted to image recognition studies \cite{carion2020end, beal2020toward}.
Recently, there has been a study on the 3D data Transformer backbone, incorporating a self-attention algorithm to enhance the receptive field. 
The main focus of these studies is to enhance the performance of learning-based detection methods by leveraging both efficient feature extraction and a multi-head self-attention modules \cite{misra2021end, mao2021voxel}. 
These techniques are aimed at improving the ability of the model to capture important spatial features and relationships within 3D data, ultimately leading to more accurate and robust detection results \cite{misra2021end, mao2021voxel}.
\par
Our idea is to apply self-attention mechanisms to sequential tasks, specifically pose prediction using LiDAR data. 
This utilization aims to capture long-range dependencies and temporal relationships, improving the accuracy of pose prediction for applications like autonomous vehicles and robotics.

\subsection{Lidar odometry}
Lidar odometry algorithms can be categorized into the following two types: geometry-based methods and learning-based methods.
The most straightforward method is point-cloud registration algorithm as known as ICP-variant \cite{besl1992method, segal2009generalized, censi2008icp} and NDT-variant \cite{biber2003normal, takeuchi20063,das2012scan, das2014scan} algorithms. 
Owing to registration algorithm's large computing consumption, feature-based LO methods \cite{shan2020lio, shan2018lego,xu2021fast} implement handcrafted feature extraction and constraint for solution space by utilizing inertial measurement unit (IMU) values. 
Geometry-based methods are intuitive for robot pose prediction and have shown high accuracy in benchmarks \cite{geiger2012we}. 
\par
Recently, CNN-based odometry methods have been proposed, and these learning-based studies have originated from research in visual odometry \cite{wang2017deepvo, zhou2017unsupervised}.
To make use of this pipeline for visual odometry, research has also been conducted on rasterizing 3D points \cite{cho2020unsupervised}.
In the wake of the PointNet proposals \cite{qi2017pointnet, qi2017pointnet++}, 3D CNN-based methods have been proposed for end-to-end registration \cite{Yew_2018_ECCV, Lu_2019_ICCV,wang2019deeppco}.
Deep point-cloud registration has been previously proposed as a solution to suppress the substantial noise that can pervade real-world data \cite{horn2020deepclr}. However, conventional geometric concepts such as the \textit{KNN} algorithm are still utilized for motion embedding part \cite{Liu_2019_CVPR}.
In addition, A learning-based registration algorithm can be extended for domain adaptation work, as it has the advantage of being able to solve localization problems without a prior map by utilizing extracted feature information \cite{cho2022openstreetmap}.
Our study builds upon the foundation provided by the open-source project referenced in \cite{deepclr}.

\section{Proposed method}
\label{sec:methods}

\subsection{Problem definition}
Input of LO problem is unordered set of 3D points $\mathcal{P} = \{\mathbf{p}_1, \dots, \mathbf{p}_{n}\}$ where $n$ is the size of the query point-cloud.
Let us define target set as $\mathcal{Q} = \{\mathbf{q}_1, \dots, \mathbf{q}_{m}\}$ where $m$ is the size of the target point-cloud.
Moreover, Individual points in the point sets are $\mathbf{p}_i = \{\mathbf{x}_i, \mathbf{f}_i\}$ and $\mathbf{q}_i = \{\mathbf{y}_i, \mathbf{g}_i\}$ where $\mathbf{x}_i, \mathbf{y}_i \in \mathbb{R}^3$ denote its a vector of position $(x, y, z)$ and $\mathbf{f}_i, \mathbf{g}_i \in \mathbb{R}^c$ are optical feature such as reflection intensity, color or surfaces where $c$ is the number of feature dimension. 
The objective of \textbf{ELiOT} is to learn $\mathbf{T}$ from the sequential set $\{\mathcal{P}, \mathcal{Q} \}$. 
Let us define $\hat{\mathbf{x}}_t \in $ SE(3) be the odometry at time $t$.
Then, we can transform $\hat{\mathbf{x}}_t$ to $\hat{\mathbf{x}}_{t+1}$ using $\mathbf{T}_t$, which denotes estimated transformation matrix at time $t$ as $\hat{\mathbf{x}}_{t+1} = \mathbf{T}_t\hat{\mathbf{x}}_t$.
Therefore, if the initial pose $\hat{\mathbf{x}}_{0}$ is obtained from $\mathbf{T}_0 = \mathbf{I}$, we can formulate the odometry pose at time $t$ as following : 
\begin{equation}
\begin{aligned}
    \hat{\mathbf{x}}_{t} = \prod_{i = 0}^{t}{\mathbf{T}_i}\mathbf{x}_{0}.
\end{aligned}
\label{eq_definition}
\end{equation}
In this study, we train our model using supervised learning, with the aim of minimizing the error between the outcome predicted by Eq. \ref{eq_definition} and the ground truth odometry. Furthermore, we utilize dual-quaternions to represent the translation and rotation components of $\mathbf{T}_t$ \cite{kenwright2012beginners, schneider2017regnet, horn2020deepclr}.

\subsection{Overall Network architecture}
\begin{figure*}[ht]
    \centering
    \includegraphics[width=2.0\columnwidth]{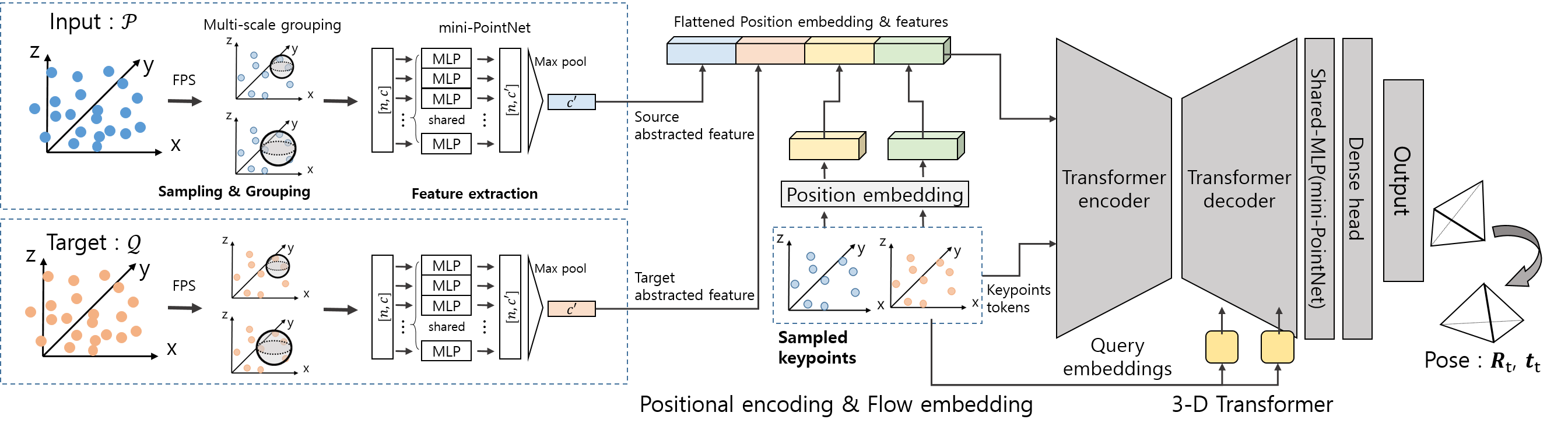}
    \caption{An overview of the proposed \textbf{ELiOT} network architecture. The process begins with a PointNet++ backbone module that extracts geometric features from two consecutive point-cloud frames, $\mathcal{P}$ and $\mathcal{Q}$. These extracted features and keypoints are then input into Implicitly Represented Flow Embedding (IRFE) layers, which learn global and local features for motion flows. Finally, the architecture utilizes Transformer-based encoder-decoder modules to predict the dual-quaternion of the consecutive point-cloud frames.}
    \label{fig:architecture}
\end{figure*}
Figure \ref{fig:architecture} presents an overview of the proposed network architecture. 
We utilize a PointNet{++} backbone module \cite{qi2017pointnet++, qi2017pointnet} in order to extract geometric features via a learning-based approach. 
This module processes two consecutive point-cloud frames, $\mathcal{P}$ and $\mathcal{Q}$, generating keypoints and their corresponding set of abstracted features.
\par
Subsequently, these keypoints and features are input into the implicitly represented flow embedding (\textit{IRFE}) layers.
This allows for the learning of global and local features for motion flows. 
Finally, Transformer-based encoder-decoder modules are employed to predict the dual-quaternion of the consecutive point-cloud frames.

\subsection{Feature extraction \& Set abstraction}
\label{subsec:sb}
The first module of our framework is designed to extract features, thereby generating a subsampled set of abstracted data.
Because the size of the $\mathbf{P}$ is too large for a training model to process directly, sampling of point clouds is necessary.
More precisely, we adopt the Furthest-Point-sampling (FPS) algorithm to extract a small number of $n_{key}$ keypoints $\mathcal{K}_{\mathcal{P}} = \{\mathbf{p}_1, \dots, \mathbf{p}_{n_{key}}\} \subset \mathbf{P}$.
As previous deep LO \cite{Liu_2019_CVPR, horn2020deepclr} extended multi-scale grouping set abstraction \cite{qi2017pointnet++} for deep registration algorithm, 
we also generate a feature set $\mathcal{F}_{\mathcal{P}} = \{\mathbf{f}_1, \dots, \mathbf{f}_{n_{key}}\}$ with $n_{r}$ radii in multi-scale grouping, where $\mathcal{F}$ is fed into PointNet \cite{qi2017pointnet} with nonlinear function $h_{sa} : \mathbb{R}^{3+c} \xrightarrow{} \mathbb{R}^{c'}$, realized as $\text{MLP}_{sa}$, and element-wise max pooling.
\par

As a result, we generate an feature aggregation set for input $\mathcal{S}^{\mathcal{P}}$ which contains $\mathcal{K}_{\mathcal{P}}$, $\mathcal{F}_{\mathcal{P}}$ and identical mechanism for target set as follows:
\begin{equation}
\begin{aligned}
    &\mathcal{S}^{\mathcal{P}}_{sa, i} = \{ \mathbf{p}_i, \mathbf{f}_i | \forall \mathbf{p}_i \in \mathcal{K}_{\mathcal{P}}, \forall \mathbf{f}_i \in \mathcal{F}_{\mathcal{P}} \}, 
    \\
    &\mathcal{S}^{\mathcal{Q}}_{sa, i} = \{ \mathbf{q}_i, \mathbf{g}_i | \forall \mathbf{q}_i \in \mathcal{K}_{\mathcal{Q}}, \forall \mathbf{g}_i \in \mathcal{F}_{\mathcal{Q}} \}.
\end{aligned}
\label{eq_sa}
\end{equation}
Then, we pass through each $\mathcal{S}^{\mathcal{P}}_{sa, i}$ and $\mathcal{S}^{\mathcal{Q}}_{sa, i}$ to batch normalization of linear layers to generate feature aggregation.
Input and target point cloud with dimensions $[n, 3+c]$ is pooled and represented with dimension $[n_{key}, 3+c']$ where $c'$ is depth of $\text{MLP}_{sa}$. 

\subsection{Implicitly represented flow embedding (\textit{IRFE})}
Our reference studies \cite{Liu_2019_CVPR, horn2020deepclr} used flow embedding modules to concatenate consecutive LiDAR frames.
Specifically, they take the sequential feature aggregation set $\mathcal{S}^{\mathcal{P}}_{sa, i}$ and $\mathcal{S}^{\mathcal{Q}}_{sa, i}$ for flow embedding module.
Using the $(x, y, z)$ position vector of $\mathbf{p}_i \in \mathcal{S}^{\mathcal{P}}_{sa, i}$ and $\mathbf{q}_i \in \mathcal{S}^{\mathcal{Q}}_{sa, i}$, they implement the $k$-nearest neighbor (\textit{KNN}) and clustering algorithms to index $n_{group}$ of corresponding points between two sequential point clouds.
Therefore, they can generate a flow embedding set $\mathcal{S}^{fe}$ by concatenate abstracted sets as follow:
\begin{equation}
\begin{aligned}
    &\mathcal{S}^{fe}_i = \{ \mathbf{q}_{i} - \mathbf{p}_{i}, \mathbf{f}_i, \mathbf{g}_i \}.
\end{aligned}
\label{eq_fe}
\end{equation}
\par
However, conventional approaches such as \textit{KNN} and clustering algorithms still have inherent limitations. 
Specifically, these methods constrain the model to correspond only with the nearest features, potentially omitting valuable distant feature interactions.
\par
To enable our model to learn uniformly distributed keypoints across the entire space, we have designed the \textit{IRFE} layer. 
This layer is intended to learn both global and local features, providing a more holistic understanding of the data.
Drawing inspiration from NeRF \cite{mildenhall2021nerf}, we employ positional encoding to transform a 3D pose into a higher dimensional space. 
This process stimulates our model to encapsulate more complex and high-frequency data about the surroundings. 
We define $\gamma(\mathbf{p})$ as the Fourier positional encoding : 
\begin{equation}
\begin{aligned}
    \gamma(\mathbf{p}) = [\mathbf{p}, &sin(2^{0}\mathbf{p}), cos(2^{0}\mathbf{p}), ..., \\
                         & sin(2^{L-1}\mathbf{p}), cos(2^{L-1}\mathbf{p}) ].
\end{aligned}
\label{eq_position_encode}
\end{equation}
where $L$ is the dimension of Transformer's decoder.
\par
Therefore, we propose the \textit{IRFE} set, denoted as $\mathcal{S}^{irfe}$. 
This set, $\mathcal{S}^{irfe}_i$, is built by concatenating the abstracted positional encoding sets with the abstracted features as follows:
\begin{equation}
\begin{aligned}
    &\mathcal{S}^{irfe}_i = \{\gamma(\mathbf{p}_i), \gamma(\mathbf{q}_i), \mathbf{f}_i, \mathbf{g}_i \},
\end{aligned}
\label{eq_irfe}
\end{equation}
where $\mathcal{S}^{irfe}_{i}$ signifies the sequence of the flattened feature set with dimensions $[n_{key}, 4 \times c']$.

\subsection{3-D Transformer-based pose prediction}
Here, we propose a encoder-decoder transformer block for self-attentioning flow embedding set.
We adopt and extend the framework of the 3D Transformer \cite{misra2021end} to perform the LO task.
\par
We feed $\mathcal{S}^{irfe}_i$, $\mathbf{p}_{i}$, and $\mathbf{q}_{i}$ into the Transformer layer. 
Motivated by the concept of learnable \textit{classification tokens} \cite{devlin2018bert, dosovitskiy2020image}, our Transformer network follows a similar approach. 
Specifically, we enable the passage of sampled keypoint tokens $\mathbf{p}_{i}$ and $\mathbf{q}_{i}$ through the Transformer layers. 
As a result, the encoder effectively processes $\mathbf{p}_{i}$ to capture spatial features, while the decoder handles $\mathbf{q}_{i}$ to focus on sequential dependencies, preserving the geometric characteristics of consecutive LiDAR frames. 
Our implementation utilizes the encoder to handle $\mathbf{p}_{i}$ and generates queries for the decoder from $\mathbf{q}_{i}$. This design choice optimizes the model's ability to accurately predict pose using LiDAR data.
\par
Finally, we employ an output multilayer perceptron using conventional Conv1d and Linear layers to predict the pose using a dual-quaternion representation. 

\subsection{Training losses}
The objective of LO is to minimize minimize the error between Eq. \ref{eq_definition} and ground truth odometry.
Here, we can denote the predictive transformation as $\mathbf{T}$, and ground truth transformation as $\mathbf{T}^{gt}$, respectively.
There is a intuitive loss function using geometric approach \cite{kendall2017geometric} such as Euler angles, SO(3) rotation matrices or quaternions.
\par
In this study, as \cite{schneider2017regnet, horn2020deepclr} proposed, we adopt dual-quaternions $\mathbf{\sigma} = \mathbf{p}_{r} + \epsilon\mathbf{q}_{d}$ for representing rigid transformation where $\mathbf{p}_{r}$, $\mathbf{q}_{d}$ are real part, dual part, respectively.
The details of dual-quaternions can be found here \cite{kenwright2012beginners}.
Therefore, we implement the dual loss as follows : 
\begin{equation}
\begin{aligned}
    &{L}_{d} = \mathbb{E}\left[\left\| \mathbf{q}_{d}^{pred} - \mathbf{q}_{d}^{gt}\right\|_2\right],
\end{aligned}
\label{eq_dual_loss}
\end{equation}
where $\mathbf{q}_{d}^{pred}$, $\mathbf{q}_{d}^{gt}$ are denoted as predicted transformation, ground-truth transformation represented by dual part in dual-quaternions form, respectively.
Also, the real loss is given by
\begin{equation}
\begin{aligned}
    &{L}_{r} = \mathbb{E}\left[\left\| \frac{\mathbf{p}^{pred}_{r}}{\left\|\mathbf{p}^{pred}_{r} \right\|} - \mathbf{p}^{gt}_{r} \right\|_2\right],
\end{aligned}
\label{eq_real_loss}
\end{equation}
where $\mathbf{p}_{r}^{pred}$, $\mathbf{p}_{r}^{gt}$ are denoted as predicted rotation, ground-truth rotation represented by real part in dual-quaternions form, respectively.

\begin{table}[t!]
\caption[]{Hyperparameters for KITTI Odometry dataset}
\label{tab:implement}
\centering
\begin{tabular}{c|c|c}
Module & Parameter for KITTI & Value \\ 
\hline 
\multirow{3}{*}{\begin{turn}{0}Feature extraction\end{turn}} 
& $n_{key}$  & 1024 \\  
& $\text{MLP}_\text{raw}$ & [[16, 16, 32], [16, 16, 32]] \\
& SA radii & [0.5, 1.0] \\
\hline
\multirow{4}{*}{\begin{turn}{0}\begin{tabular}[c]{@{}c@{}}Flow embedding \& \\ Positional encoding \end{tabular}\end{turn}}
& $L$ & 64 \\
& PE method & Fourier \\
& PE normalization & True \\
& Gauss scale & 1.0 \\
\hline
\multirow{13}{*}{\begin{turn}{0}Transformer\end{turn}}
& Enc. layer & 3 \\
& Enc. dim. & 256 \\
& Enc. head number & 4 \\
& Enc. ffn dim.& 16 \\
& Enc. dropout& 0.1 \\
& Enc. activation & relu \\
& Dec. layer & 3 \\
& Dec. dim. & 256 \\
& Dec. head number & 2 \\
& Dec. ffn dim.& 16 \\
& Dec. dropout& 0.1 \\
& Num. of queries & 1024 \\
\hline
\multirow{2}{*}{\begin{turn}{0}Head\end{turn}}
& $\text{MLP}_\text{pn}$ & [256, 512, 1024] \\
& $\text{MLP}_\text{fc}$ & [1024, 512, 256, 8] \\
\hline
\end{tabular}
\end{table} 
\section{Implementation details}
\label{sec:implement}
In our implementation, raw points are consecutively sampled using FPS methods, and aggregated into feature abstracted set $\mathcal{S}_{sa, i}^{\mathcal{P}}$ and $\mathcal{S}_{sa, i}^{\mathcal{Q}}$.
\par
When it comes to the flow embedding module, we opt for a departure from conventional methods such as the \textit{KNN} algorithm. 
Instead, we have opted for an \textit{IRFE} layer, which involves flattening $\gamma(\mathbf{p}_i)$, $\gamma(\mathbf{q}_i)$, $\mathbf{f}_i$, and $\mathbf{g}_i$, each having a consistent dimension of $[n_{key}, c']$. 
This flattening process yields a sequence of the flattened feature set with dimensions of $[n_{key}, 4 \times c']$.
\par
Lastly, our transformer block is composed of position embedding, encoder, decoder, and query embedding modules.
An overview of the network parameter is represented in Table \ref{tab:implement}.
\par
During learning process, random translations and rotations noises are applied to the input points set for data augmentation.

\begin{table*}[htb!]
\caption[]{Comparison with the state-of-the-arts on KITTI odometry dataset. We list the classic methods and learning-based odometry methods.}
\label{tab:result}
\centering
\resizebox{0.9\textwidth}{!}{%
\begin{tabular}{c|c|cccccccccc}
\multirow{2}{*}{Method} & \multirow{2}{*}{Seq.} 
& \multicolumn{2}{|c}{7} & \multicolumn{2}{|c}{8} & \multicolumn{2}{|c}{9} & \multicolumn{2}{|c}{10} & \multicolumn{2}{|c}{Testing Avg.} \\ \cline{3-12}
& & \multicolumn{1}{|c}{$t_{rel}$} & \multicolumn{1}{|c}{$r_{rel}$} & \multicolumn{1}{|c}{$t_{rel}$} & \multicolumn{1}{|c}{$r_{rel}$} & \multicolumn{1}{|c}{$t_{rel}$} & \multicolumn{1}{|c}{$r_{rel}$} & \multicolumn{1}{|c}{$t_{rel}$} & \multicolumn{1}{|c}{$r_{rel}$} & \multicolumn{1}{|c}{$t_{rel}$} & \multicolumn{1}{|c}{$r_{rel}$} \\
\hline \hline 
\multirow{4}{*}{\begin{turn}{90}Classic\end{turn}}
& ICP-PO2PO & $5.17$ & $3.35$ & $10.04$ & $4.93$ & $6.93$ & $2.89$ & $8.91$ & $4.74$ & $7.76$ & $3.98$\\
& ICP-PO2PL & $1.55$ & $1.42$ & $4.42$ & $2.14$ & $3.95$ & $1.71$ & $6.13$ & $2.60$ & $4.01$ & $1.97$\\
& GICP\cite{segal2009generalized}  & $\textbf{0.64}$ & $\textbf{0.45}$ & $\textbf{1.58}$ & $\textbf{0.75}$ & $\textbf{1.97}$ & $\textbf{0.77}$ & $\textbf{1.31}$ & $\textbf{0.62}$ & $\textbf{1.38}$ & $\textbf{0.65}$\\
& LOAM\cite{zhang2014loam}  & $2.98$ & $1.55$ & $4.89$ & $2.04$ & $6.04$ & $1.79$ & $3.65$ & $1.55$ & $4.39$ & $1.73$\\
\hline
\multirow{4}{*}{\begin{turn}{90}\begin{tabular}[c]{@{}c@{}} Learning-\\ based \end{tabular} \end{turn}}
& SelfVoxeLO\cite{xu2021selfvoxelo} & $3.09$ & $1.81$ & $3.16$ & $1.14$ & $3.01$ & $1.14$ & $3.48$ & $1.11$ & $3.19$ & $1.30$\\
& Efficient-LO\cite{wang2021efficient} & $\textbf{0.46}$ & $\textbf{0.38}$ & $\textbf{1.14}$ & $\textbf{0.41}$ & $\textbf{0.78}$ & $\textbf{0.33}$ & $\textbf{0.80}$ & $\textbf{0.46}$ & $\textbf{0.80}$ & $\textbf{0.40}$\\
& DEEPCLR\cite{horn2020deepclr} & $4.82$ & $3.47$ & $6.37$ & $2.27$ & $3.87$ & $1.63$ & $8.10$ & $2.77$ & $5.79$ & $2.54$\\
& \textbf{Proposed} & $\textbf{7.20}$ & $\textbf{3.39}$ & $\textbf{5.30}$ & $\textbf{1.85}$ & $\textbf{5.42}$ & $\textbf{2.09}$ & $\textbf{12.46}$ & $\textbf{3.35}$ & $\textbf{7.59}$ & $\textbf{2.67}$\\
\hline
\end{tabular}
 }
\begin{flushleft}
* Units for $t_{rel}$ and $r_{rel}$ are the average translational RMSE (\%) and rotational RMSE ($\deg$/100m) respectively on all possible subsequences in the length of 100, 200, ..., 800 m.
\end{flushleft}
\end{table*}

\begin{figure*}[!hbt]
    \centering
    \subfigure[Sequence 07]{
        \includegraphics[width=0.8\columnwidth]{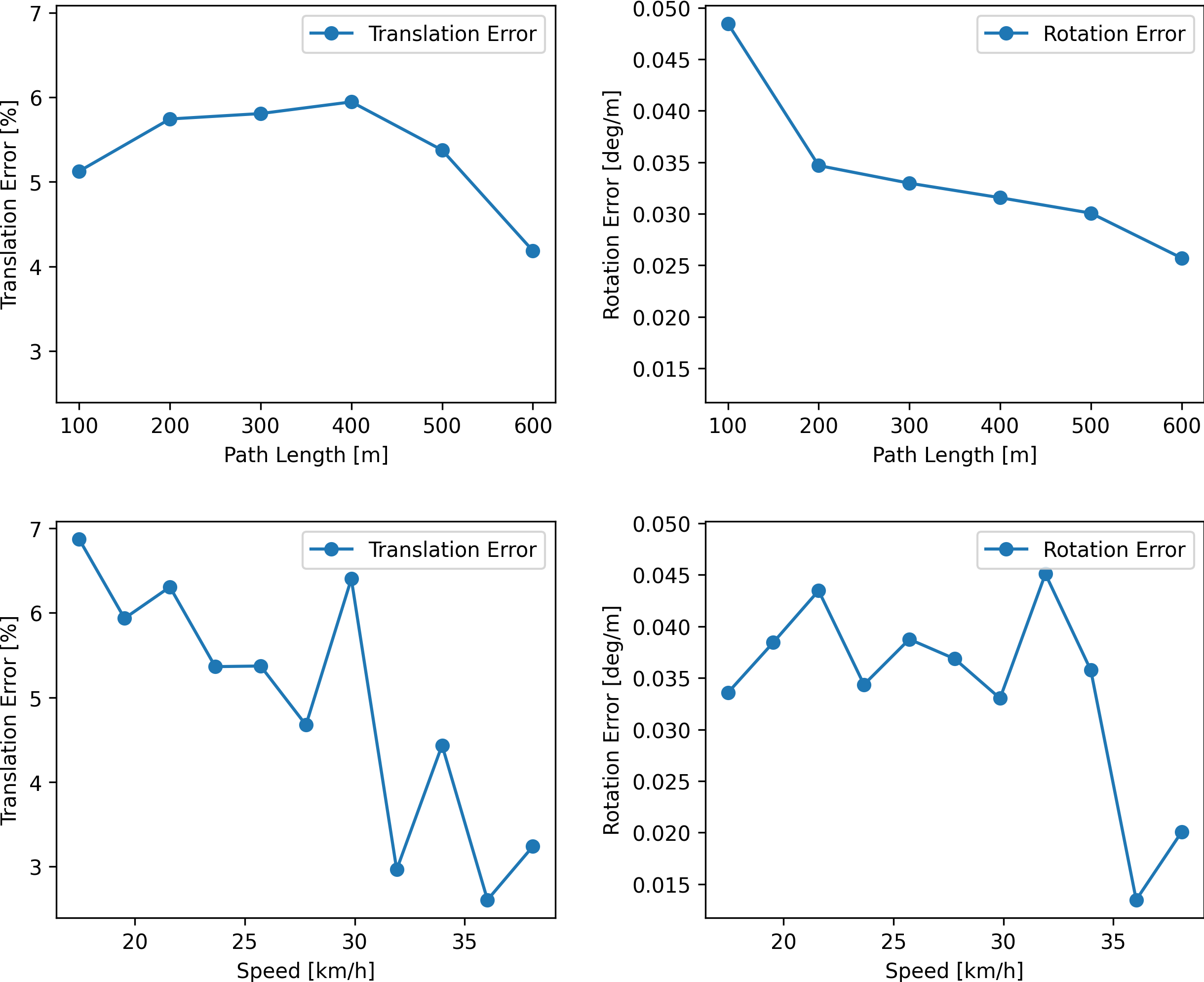}
        \label{fig:07error}
    }
    \subfigure[Sequence 08]{
        \includegraphics[width=0.8\columnwidth]{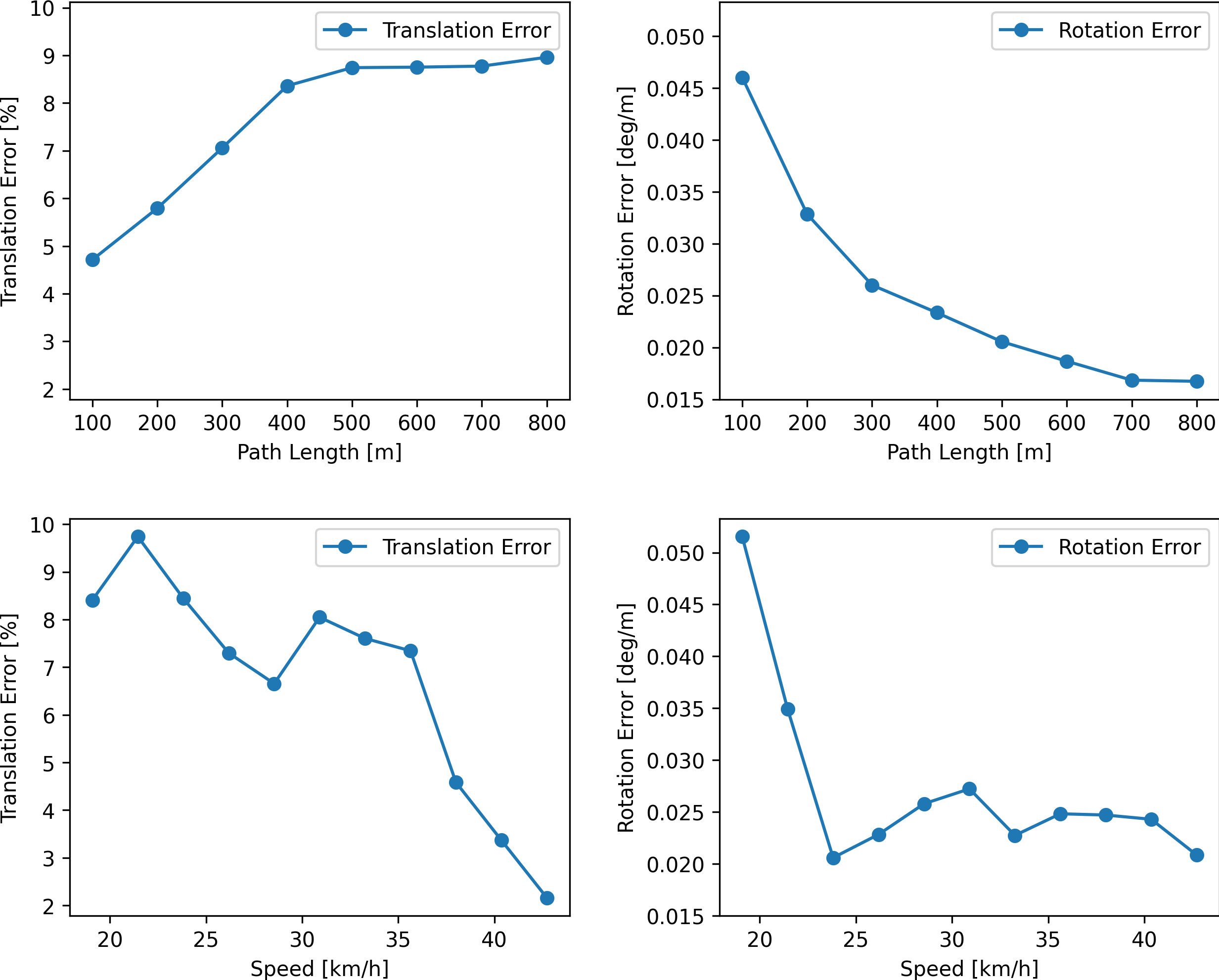}
        \label{fig:08error}
    }
    \subfigure[Sequence 09]{
        \includegraphics[width=0.8\columnwidth]{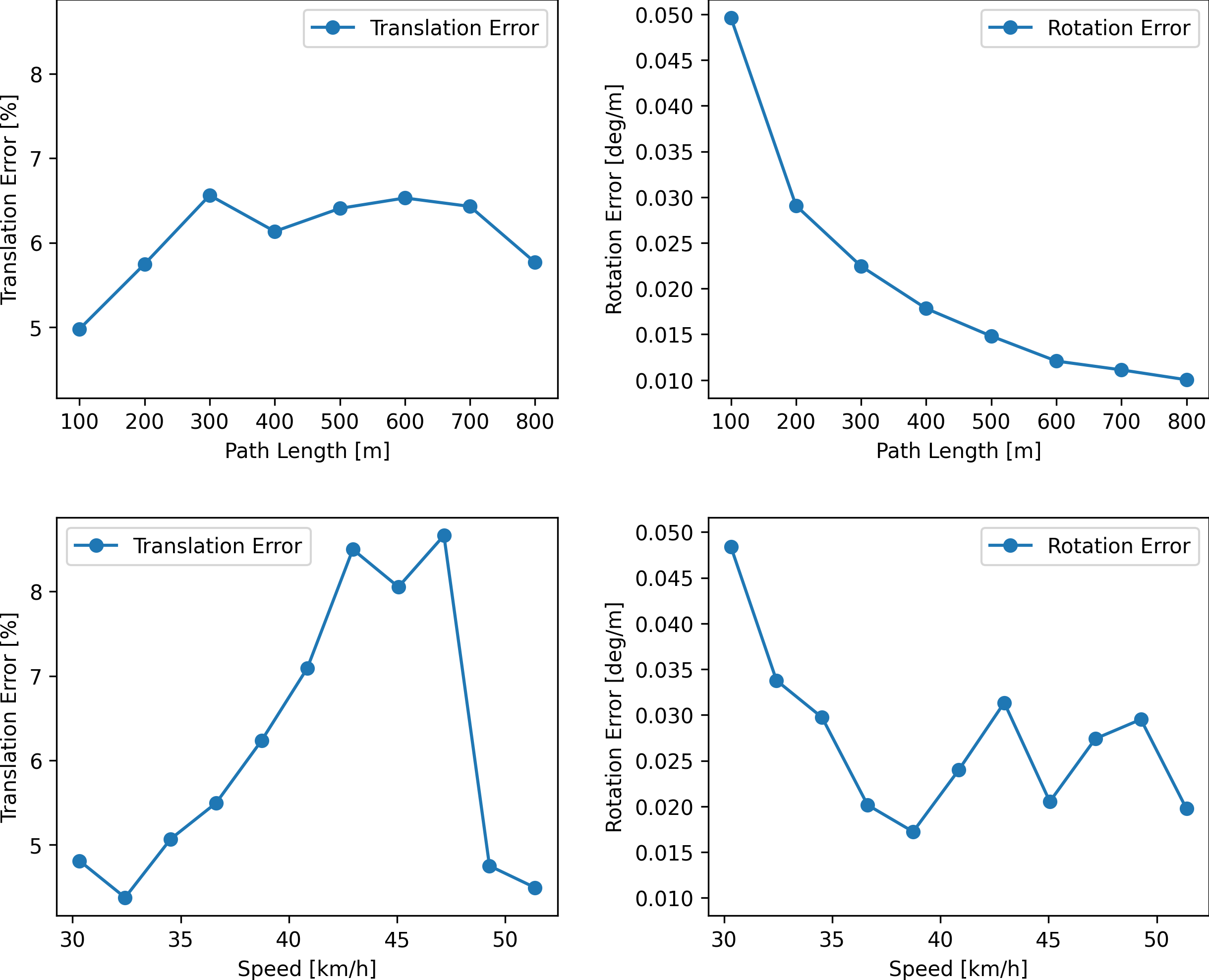}
        \label{fig:09error}
    }
    \subfigure[Sequence 10]{
        \includegraphics[width=0.8\columnwidth]{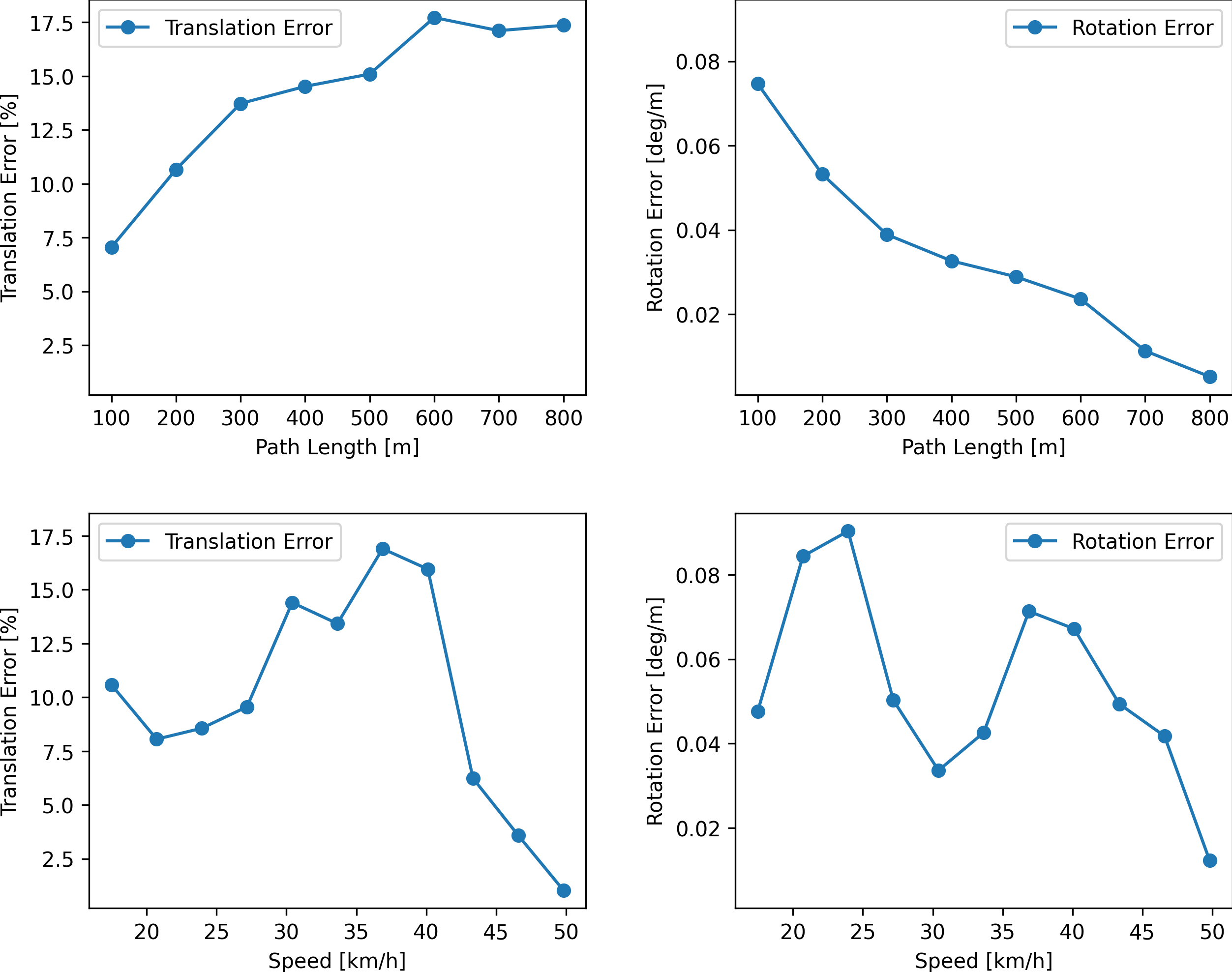}
        \label{fig:10error}
    }
    \caption{Illustration of translation and rotation errors on KITTI odometry sequences 07 to 10. Each subfigure demonstrates the respective errors for a particular sequence, showcasing the performance of our proposed method according to segment progress and up to speed.}
    \label{fig:kitti_error}
\end{figure*}

\begin{figure*}[!hbt]
    \centering
    \includegraphics[width=1.9\columnwidth]{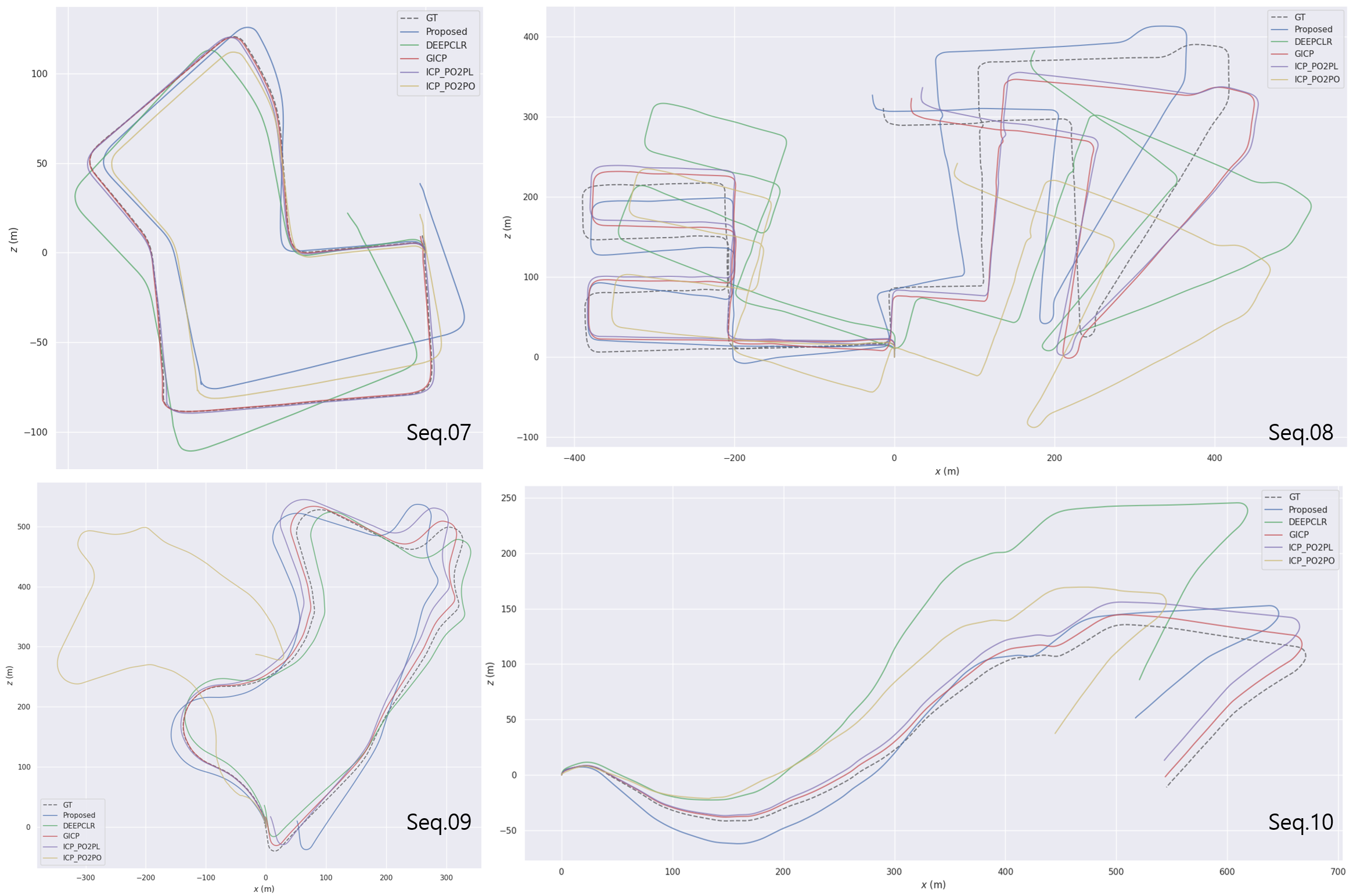}
    \caption{Trajectories generated by the proposed ELiOT model on the KITTI odometry dataset. Each subfigure showcases the effectiveness of the proposed approach in diverse driving conditions and environments.}
    \label{fig:kitti_07_10}
\end{figure*}

\begin{figure*}[ht]
    \centering
    \includegraphics[width=1.9\columnwidth]{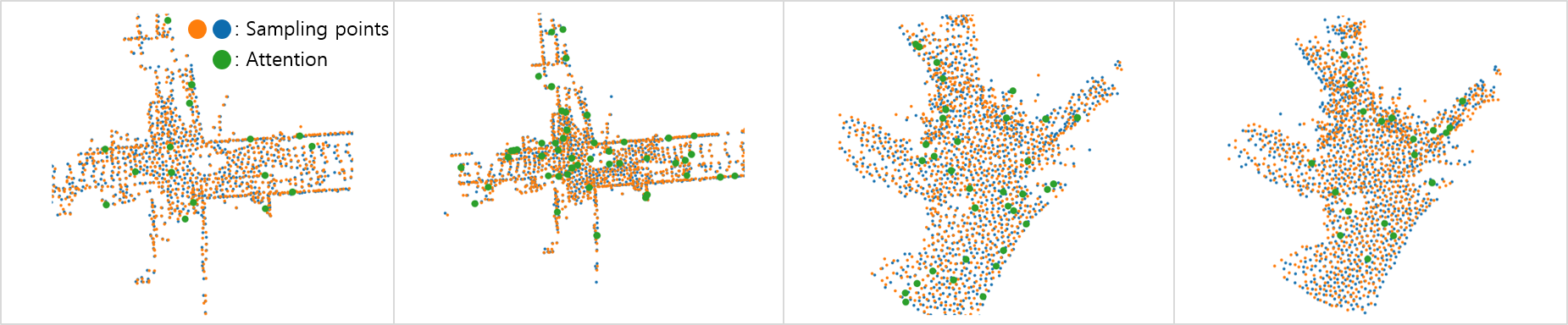}
    \caption{Visualizing flow attention results: Emphasizing edge and corner features for enhanced motion information in LiDAR frames}
    \label{fig:attention}
\end{figure*}

\section{Experiments}
\label{sec:experiment}
\subsection{KITTI Odometry}
We validate the proposed end-to-end LO method with KITTI odometry dataset \cite{geiger2012we}.
There are 22 LiDAR sequences and corresponding RGB/gray images in the KITTI odometry dataset.
It provides ground-truth poses derived from IMU/GPS fusion algorithms for sequences 00-10.
Also, ground-truth poses aren't provided by the remaining sequences, which are for benchmark testing.
There are different types of road environments in this dataset, as well as pedestrians, cyclists, and different types of vehicles. 
Data-collection vehicle drives from 0 $km/h$ to $90km/h$ in different areas.
Therefore, we can evaluate whether the trained model can handle noisy real-world data with a large-scale point cloud.

\subsection{Evaluation}
All methods are tested on a machine equipped with an Intel Core i9-10900X CPU @ 3.70GHz, memory of 128GB, and a GPU of 24GB.
The implementation is written in PyTorch.
\par
As our training dataset consists of seqeunces 00-06, our method is compared against other competitive methods using sequences 07-10 from the KITTI odometry dataset. 
We compare our method with the classic method \cite{censi2008icp, segal2009generalized} ---i.e. to demonstrate the classic methods, point-to-point ICP, point-to-plane ICP, and GICP are implemented based on the Open3D library.   
Additionally, we compare the proposed method against other CNN-based odometry methods \cite{horn2020deepclr, wang2019deeppco}, and we discover that our method achieves competitive performance.
We train a model with 00-06 sequences and evaluate with 07-10. 
The results were evaluated using the \textit{KITTI Odometry Devkit}, which provides the average relative translation ($t_{rmse}$) and rotation errors ($r_{rmse}$) as part of its output metrics.
\par
The evaluation results, which include a comparison with classic and learning-based methods, are presented in Table \ref{tab:result}. 
For a more intuitive understanding of the performance of our proposed method, we have visualized the generated trajectories on KITTI odometry sequences 07 through 10, as depicted in Fig. \ref{fig:kitti_07_10}. This visualization also includes each sequence's translation and rotation error segregated by speed and path length segments in Fig. \ref{fig:kitti_error}.
\par
While comparing our method with traditional techniques and referencing learning-based studies, we acknowledge the potential for further enhancements. 
Specifically, we understand that our methodology requires more development to achieve the performance benchmarks set by 3D to 2D projection-based techniques and conventional approaches. 
As we continue our research, we aim to address these areas of improvement and strive for better results in future iterations.
\par
One possible reason for the lack of generalization could be that the KITTI dataset used for training only covered limited sequences from 0 to 6. 
As a result, our model may not have fully generalized to other unseen sequences. Addressing this limitation will be an important focus for future work.

\subsection{Visualization}
In Fig. \ref{fig:attention}, we provide a visual representation of the sampling points and their respective flow attention results. 
These results highlight how the attention is primarily directed towards edge and corner features. 
These features provide a richer source of motion information between consecutive LiDAR frames, thereby significantly contributing to the efficacy of our method.

\section{Conclusion}
\label{sec:conclusion}
In this work, we presented \textbf{ELiOT}, a novel architecture that seamlessly integrates point feature abstraction with a motion flow self-attentioning 3D transformer framework. 
\par
While comparing our method with traditional techniques and referencing learning-based studies, we acknowledge the potential for further enhancements. We understand that our methodology requires more development to achieve the performance benchmarks set by 3D to 2D projection-based techniques and conventional approaches.
\par
Nonetheless, our deep learning-based LiDAR odometry method holds significant promise. Leveraging the potential of positional embedding networks to identify motion flow and emphasizing key features, our approach achieves geometric motion through an end-to-end framework. This makes it a valuable contribution to the field, offering a unique and promising perspective for advancements in autonomous navigation systems.
\par
To address the generalization issue, we plan to expand our training dataset with diverse sequences. Additionally, exploring the integration of novel techniques and architectural enhancements will further improve \textbf{ELiOT}'s capabilities and push the boundaries of LiDAR odometry accuracy and efficiency.



\bibliographystyle{unsrt}
\bibliography{cite}


\end{document}